\documentclass[twocolumn]{aastex61}
\usepackage{graphicx}

\newcommand{\eg}{\mbox{e.g.}}

\newcommand{\as}{\mbox{$^{\prime\prime}$}}

\newcommand{\kms}{\mbox{\,km\,s$^{-1}$}}
\newcommand{\kmsec}{\mbox{\,km\,s$^{-1}$}}

\newcommand{\um}{\,$\mu$m}

\newcommand{\av}{\mbox{$\rm A_V$}}

\newcommand{\vlsr}{\mbox{$\rm v_{LSR}$}}
\newcommand{\teff}{\mbox{T$_{\rm eff}$}}
\newcommand{\logg}{\mbox{log~{\it g}}}
\def\vmicro{\mbox{$\xi_{\rm t}$}}
 
\newcommand{\hh}{H$_2$}

\newcommand{\nco}{$\rm N_{CO}$}
\newcommand{\bco}{$\rm b_{CO}$}
\newcommand{\nhh}{$\rm N_{H_2}$}
\newcommand{\hhco}{$\rm N_{H_2}$/$\rm N_{CO}$} 
\newcommand{\hhav}{$\rm N_{H_2}$/$\rm A_V$}
\newcommand{\coav}{$\rm N_{CO}$/$\rm A_V$}

\received{}
\revised{}
\accepted{}
\submitjournal{ApJ}

\shorttitle{\hh\ and CO Absorption}
\shortauthors{Lacy et al.}

\begin{document}


\title{ \hh , CO, and Dust Absorption Through Cold Molecular Clouds}

\email{lacy@astro.as.utexas.edu, chris@astro.as.utexas.edu, hkim@gemini.edu, dtj@astro.as.utexas.edu}

\author{John H. Lacy}
\affil{Department of Astronomy and McDonald Observatory, University of Texas, Austin, TX 78712}
\author{Christopher Sneden}
\affil{Department of Astronomy and McDonald Observatory, University of Texas, Austin, TX 78712}
\author{Hwihyun Kim}
\affil{Department of Astronomy and McDonald Observatory, University of Texas, Austin, TX 78712}
\affil{Korea Astronomy and Space Science Institute, Daejeon, Republic of Korea.}
\affil{Current address: Gemini Observatory, c/o AURA, Casilla 603, La Serena, Chile.}
\author{Daniel T. Jaffe}
\affil{Department of Astronomy and McDonald Observatory, University of Texas, Austin, TX 78712}

\begin{abstract}

The abundance of \hh\ in molecular clouds, relative to the commonly used tracer CO, has only been measured toward a few embedded stars, which may be surrounded by atypical gas.
We present observations of near-infrared absorption by \hh , CO, and dust toward stars behind molecular clouds, providing a representative sample of these molecules in cold molecular gas,
primarily in the Taurus Molecular Cloud.
We find \hhav\ $\rm \approx 1.0 \times 10^{21} cm^{-2}$, \coav\ $\rm \approx 1.5 \times 10^{17} cm^{-2}$ ($1.8 \times 10^{17}$ including solid CO), and \hhco\ $\approx 6000$.
The measured \hhco\ ratio is consistent with that toward embedded stars in various molecular clouds, but both are less than that derived from mm-wave observations of CO and star counts.
The difference apparently results from the higher directly measured \coav\ ratio.

\end{abstract}





\section{Introduction}

Molecular hydrogen, \hh , is undoubtedly the most abundant species in molecular clouds.
It and atomic helium must constitute $\sim$98\% of the mass in molecular clouds.
However, both of these species are very difficult to observe.
Cold helium gas has no detectable transitions at wavelengths that are observable through molecular clouds.
\hh\ has electric quadrupolar vibrational and rotational transitions in the infrared, but the weakness of these lines makes them very difficult to observe in absorption, and the wide spacing between the \hh\ energy levels causes their emission lines to be observable only from unusually hot or radiatively excited gas.

Due to the difficulty of observing \hh , CO is most often used to measure the column density of gas in molecular clouds, with an assumed \hh /CO abundance or line flux ratio.
A number of authors have attempted to determine these ratios in order to calibrate cloud masses determined from CO observations.
\citet{dickman78} made millimeter wavelength J=1-0 observations of $^{13}$CO, which he compared to determinations of \av\ based on star counts.
Assuming \hhav\ = $\rm 1.25\times 10^{21} cm^{-2}$, he derived \nhh\ = $\rm 5 \times 10^5\, N_{^{13}CO}$.
Assuming $\rm ^{12}CO / ^{13}CO$ = 60, this corresponds to \hhco\ = 8300.
\citet{FLW82} extended Dickman's work to larger column densities and rarer CO isotopomers, which are less affected by line saturation.
They found somewhat different relations for the Taurus and Ophiuchus molecular clouds and for different CO isotopomers.
Their results are generally consistent with $\rm N_{C^{18}O} / A_V = 1.7 \times 10^{14} cm^{-2}$, or \hhco\ = $1.2 \times 10^4$ for the interiors of clouds, at $\rm A_V > 4$.
They also give the ratio known as $\rm X_{CO} = N_{H_2} / I_{CO} = 1.8\times 10^{20} cm^{-2} (K\,km\,s^{-1})^{-1}$, which is used to derive \hh\ column densities from CO observations.
More recently, \citet{pineda10} measured \nco\ and \av\ at higher angular resolution in the Taurus Molecular Cloud.
They found a non-linear relation between \nco\ and \av\ consistent with freeze-out of CO at \av\ $> 10$, and a variation in the \coav\ ratio with location in the cloud.
On average they found $\rm N_{CO} / A_V = 1.0 \times 10^{17} cm^{-2}$ at \av\ $< 10$, and assuming \hhav\ = $9.4 \times 10^{20}$ cm$^{-2}$, \hhco\ = 9000.
Other recent studies include \citet{pineda08} and \citet{lee14} who measured $\rm X_{CO}$ in the Perseus molecular cloud.
\citet{lee14} found substantial variations in $\rm X_{CO}$ among regions in Perseus, and large variations in $\rm X_{CO}$ on small scales.
These and other determinations of \hhco\ and $\rm X_{CO}$ are reviewed by \citet{bolatto13}.
However, most determinations of the relation between the \hh\ column density and the CO column density or emission depend on the assumed \hhav\ ratio, which might not be valid in dense molecular clouds.

Although infrared lines of \hh\ are quite weak, it is possible to observe them in absorption through sufficiently large columns of molecular gas.
The 2-2.4\um\ v=1-0 lines are the most favorable, as they are the strongest infrared lines of \hh.
They also have the advantages that they lie close in wavelength to the v=2-0 band of CO, allowing simultaneous, or nearly simultaneous measurements of column densities of these two molecules, and that they lie in a relatively transparent region of the telluric spectrum, in the near-infrared K band.
There have been three direct determinations of \hhco\ in molecular clouds using the K-band absorption lines of these molecules.
\citet{lacy94} detected absorption in the v=1-0, J=2-0 (or S(0)) line of \hh\ and low-J lines of the v=2-0 band of CO toward the embedded star NGC 2024 IRS 2.
They derived an \hhco\ ratio of $4000 \pm 3000$.
\citet{kulesa02} observed these lines toward five additional high mass stars embedded in molecular clouds.
His observations are consistent with \hhco\ = 3500-7000, with all measured ratios being less than the ratios determined by \citet{dickman78} and \citet{FLW82}.
\citet{goto15} added one additional source, NGC 7538 IRS 1, and measured \hhco\ = 3600.

However, all of the sources in which \hh\ and CO infrared absorption has been observed are embedded stars, and from the relatively high temperatures derived from the CO lines most of the absorption must occur in gas close to the stars, which may have unusual chemistry.
It is also possible that UV radiation from the stars excites \hh\ emission, which could contaminate the absorption lines.
In this paper we present observations of stars lying behind molecular clouds.
The lines of sight to these stars should probe more typical molecular cloud gas and should be uncontaminated by \hh\ emission; even if there is diffuse emission along the lines of sight to these stars it would be removed by the sky subtraction during data reduction.

\section{Observations}

Observations were made with the Immersion Grating Infrared Spectrograph (IGRINS) \citep{park14, mace16} on the 2.7m Harlan J. Smith telescope at McDonald Observatory.
IGRINS is a high resolution (R = 45,000) near-infrared cross-dispersed spectrograph, which simultaneously covers the H and K spectral bands (1.45\um\ - 2.45\um ).
For this project only the K band, which includes the v=1-0 S(0), S(1), and Q-branch lines of \hh\ and the v=2-0 P and R branches of CO, was used.
The telescope was nodded to move object images between two positions along the 1\as $\times$15\as\ entrance slit in an ABBA pattern.
The exposure time per nod position was 300 sec, and typically six ABBA sequences, or two hours on-source time was acquired for each target, with a goal of a statistical S/N of 1000.
Nearby A0V stars were used for telluric comparison, with similar exposure times and somewhat better S/N ratios.

Three categories of targets were observed: young stars in molecular clouds still surrounded by their natal gas and dust, which we refer to as embedded stars; stars lying behind molecular clouds, which we refer to as background stars; and relatively unextincted stars that were observed to test our photospheric spectrum models, which we refer to as foreground stars.
Four embedded stars in the Taurus and Monoceros clouds were observed.
Eight background stars were observed, five lying behind the Taurus cloud and three lying behind clouds in Ophiuchus and Serpens.  All but one of these were late-type giants, with complicated photospheric spectra, necessitating modeling to extract interstellar lines, as is described below.
And five foreground stars, with similar spectral types to the background giants, were observed.
Observations of several other stars were attempted, which either had too little extinction to show interstellar absorption or too low S/N to provide useful results.
A list of targets and observing parameters is given in Table 1.

\begin{deluxetable*}{ccccccc}
\tablecaption{Target and observation parameters}
\tablewidth{0pt}
\tablehead{
\colhead{Name} & \colhead{Sp type\tablenotemark{a}} & \colhead{location} & \colhead{K mag} &
\colhead{dates\tablenotemark{b}} & \colhead{int time}
}
\startdata
embedded stars &&&&&\\
Elias~3-1 & YSO & Taurus & 5.8 & 2015/12/03 & 40m \\
Elias~3-18 & YSO & Taurus & 6.3 & 2015/12/03 & 60m \\
IRAS 04278+2253 & YSO & Taurus & 5.8 & 2015/12/03 & 40m \\
AFGL~989\tablenotemark{c} & Cl II YSO & Monoceros & 4.9 & 2015/12/03 & 40m \\
background stars &&&&&\\
Elias~3-13 & K2 III & Taurus & 5.5 & 2015/12/02,04 & 40+80m \\
Elias~3-16 & K2 III & Taurus & 5.2 & 2015/12/02,04 & 40+80m \\
Tamura~8 & K5 III & Taurus & 7.5 & 2015/12/02,04 & 80+40m \\
HD~283809 & F8? & Taurus & 6.2 & 2016/01/23,24,25 & 40+40+40m \\
Kim~1-59 & K2 III & Taurus & 7.9 & 2015/12/03 & 40m \\
Elias~2-15 & M6 III & Ophiuchus & 5.3 & 2015/06/18 & 40m \\
Elias~2-35 & K6 III & Ophiuchus & 7.3 & 2015/07/24 & 60m \\
SVS76 Ser 9 & G0 III? & Serpens & 8.5 & 2015/07/24 & 80m \\
foreground stars &&&&&\\
HR 5899 & K4 III & Serpens & 2.3 & 2016/07/17 & 8m \\
HR 7800 & K7 III & Cygnus & 2.0 & 2016/07/15 & 12m \\
HR 7919 & K2 III & Cygnus & 3.3 & 2016/07/15 & 12m \\
HR 7956 & K3 III & Cygnus & 1.8 & 2016/07/17 & 8m \\
HR 7969 & K5 III & Cygnus & 2.3 & 2016/07/17 & 12m \\
\enddata
\tablenotetext{a}{Spectral types taken from SIMBAD; generally uncertain.}
\tablenotetext{b}{Precipitable water vapor was $\sim$30 mm for 2015/06-07, $\sim$2 mm for 2015/12,  $\sim$3 mm for 2016/01, and $\sim$30 mm for 2016/07 observations.}
\tablenotetext{c}{AFGL 989 = NGC 2264 IRS 1 = Allen's Star}
\end{deluxetable*}

The data were reduced with a custom fortran pipeline, which performed standard procedures of spike removal, flat-fielding with a lamp spectrum, distortion correction, and point-source spectrum extraction.
Telluric absorption lines in comparison star spectra were used for wavelength calibration.
After spectrum extraction, the target spectra were divided by comparison source spectra, with small spectral shifts and corrections for airmass differences introduced to minimize telluric residuals.
In addition, residuals due to changes in the telluric water vapor between the target and comparison star observations were corrected by adding or subtracting a small fraction of a model for the telluric water absorption.

\begin{figure}
\includegraphics[angle=0,scale=0.50]{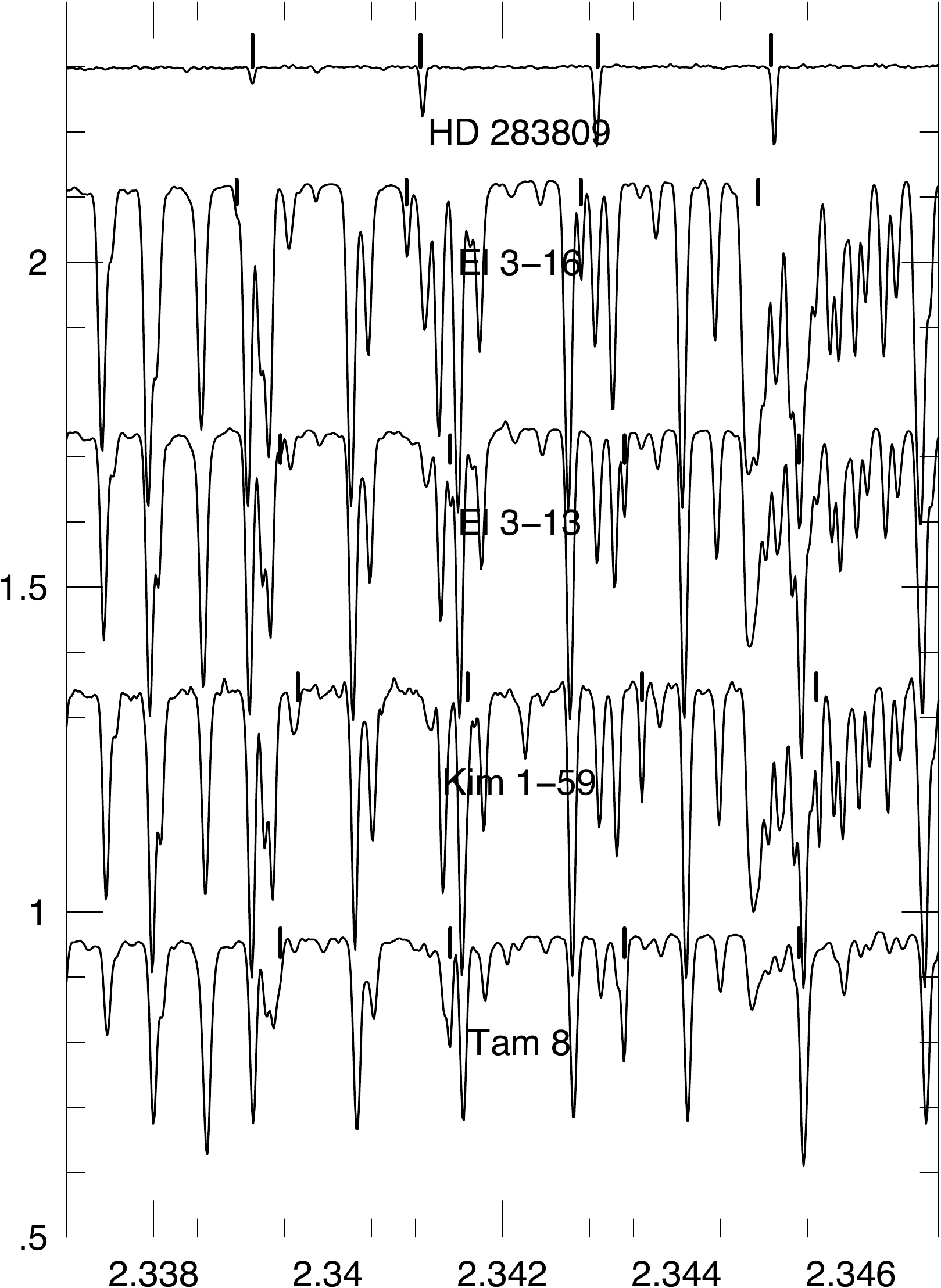}
\caption{Spectra of stars behind the Taurus Molecular Cloud in the CO v=2-0 band region.
Spectra (except HD~283809) have been shifted to the stellar rest frames.
Shifted positions of the interstellar CO R(3)--R(0) lines are marked with ticks at v$_{LSR}$ = +7 \kms.
Spectra are normalized to a continuum level of 1.0 and offset.
All figures are corrected for telluric absorption.
\label{COTaufig}}
\end{figure}

\begin{figure}
\includegraphics[angle=0,scale=0.45]{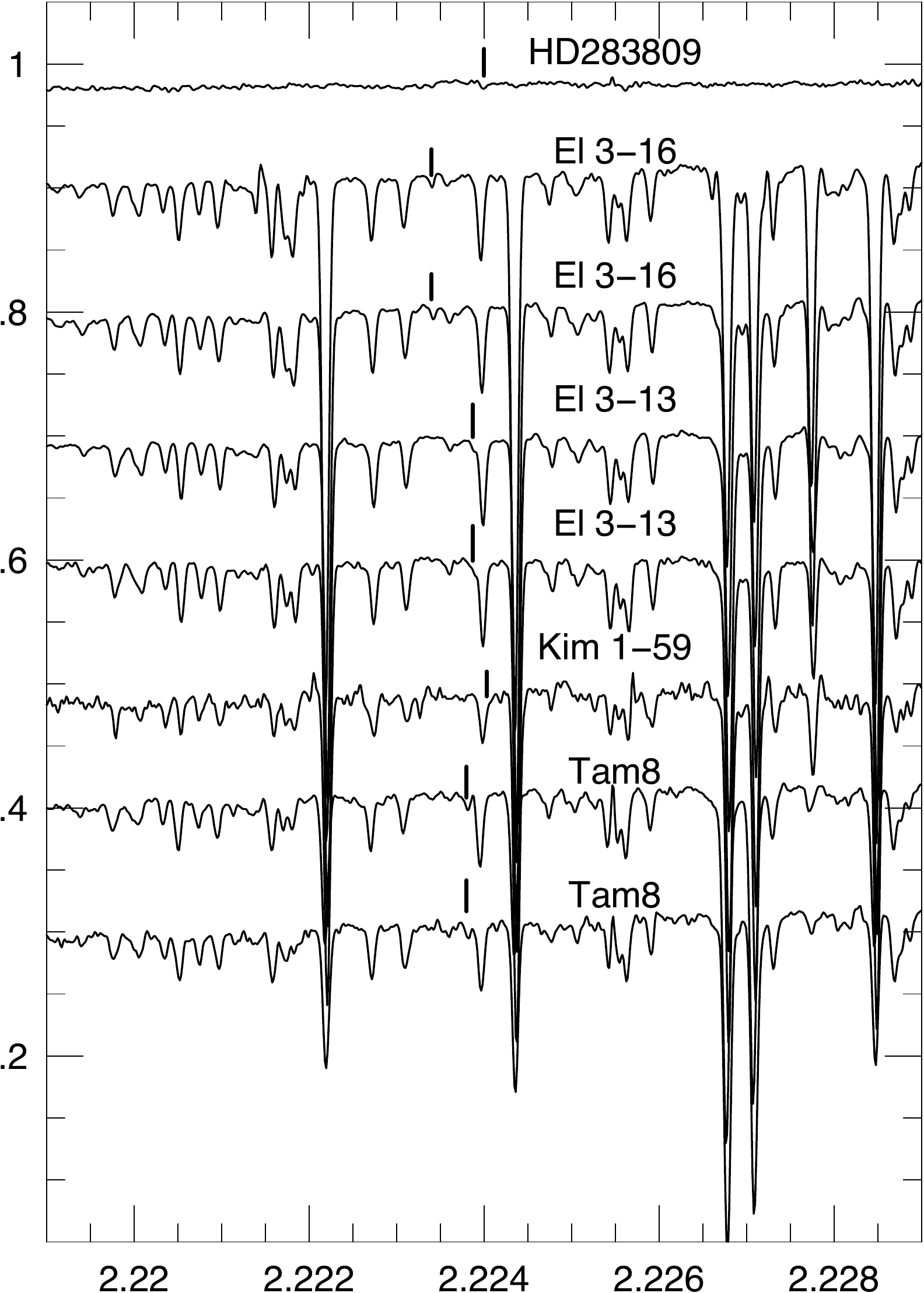}
\caption{Spectra of stars behind the Taurus Molecular Cloud in the \hh\ v=1-0 S(0) line region.
Spectra (except HD~283809) have been shifted to the stellar rest frames.
Shifted positions of the S(0) line are marked with ticks at the CO band v$_{LSR}$ of +7 \kms .
Spectra observed on different nights are shown separately to allow an estimate of the noise level.
Other figures show time-weighted averages of spectra from multiple nights.
\label{H2Taufig}}
\end{figure}

Spectra of the Taurus background sources are shown in Figure~\ref{COTaufig} (low-J CO band region) and Figure~\ref{H2Taufig} (\hh\ S(0) line region).
The spectra are shifted to the stellar rest frames, with shifted positions of the interstellar lines marked.
Wavelengths are in vacuum.
In Figure~\ref{H2Taufig} observations from different nights are shown separately to allow an estimate of the noise level.
All other figures show time-weighted averages of data from multiple nights.
All plotted spectra are corrected for telluric absorption.
Although photospheric features are generally much stronger than the interstellar lines, comparison of the spectra makes it apparent that \hh\ absorption was detected toward Elias~3-16 and Tamura~8.
\hh\ absorption is probably present in Elias~3-13 and Kim~1-59, but it is blended with a photospheric line.
HD~283809 is a hot star, lacking strong photospheric features, but with less foreground absorption, making the detection of \hh\ absorption less certain, although a feature is present at the expected wavelength.
The interstellar CO lines are more difficult to recognize in the observed spectra due to the strong photospheric CO lines, but are apparent after division by models, as discussed below.
No sources show evidence of \hh\ S(1) absorption.

\begin{figure}
\includegraphics[angle=270,scale=0.40]{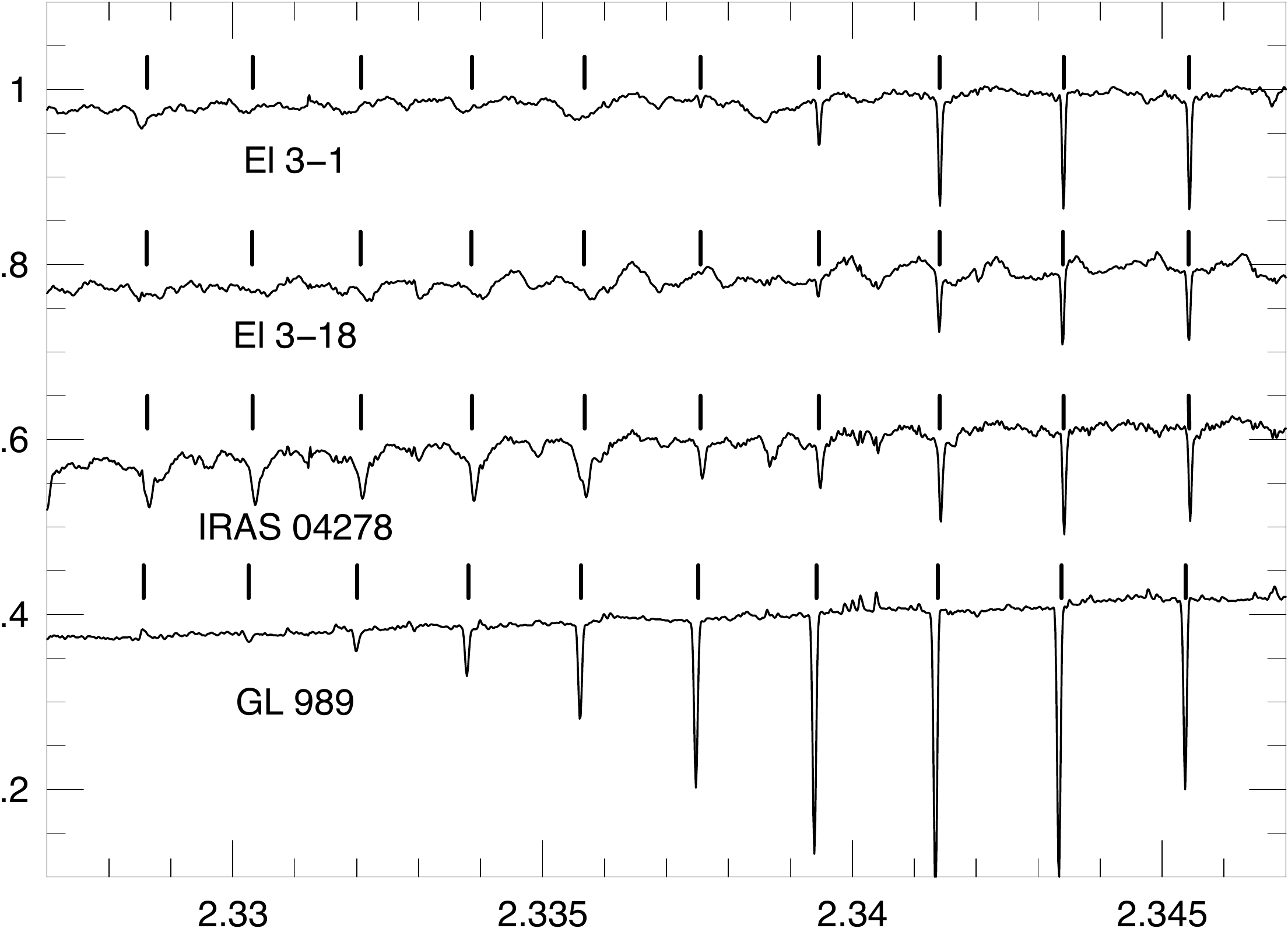}
\caption{Spectra of stars embedded in the Taurus and Monoceros molecular clouds in the CO R(9) -- R(0) region.
Positions of the interstellar CO lines are marked with ticks at v$_{LSR}$ = +7 \kms\ for the Taurus sources and +9 \kms\ for GL 989.
\label{COembdfig}}
\end{figure}

\begin{figure}
\includegraphics[angle=270,scale=0.40]{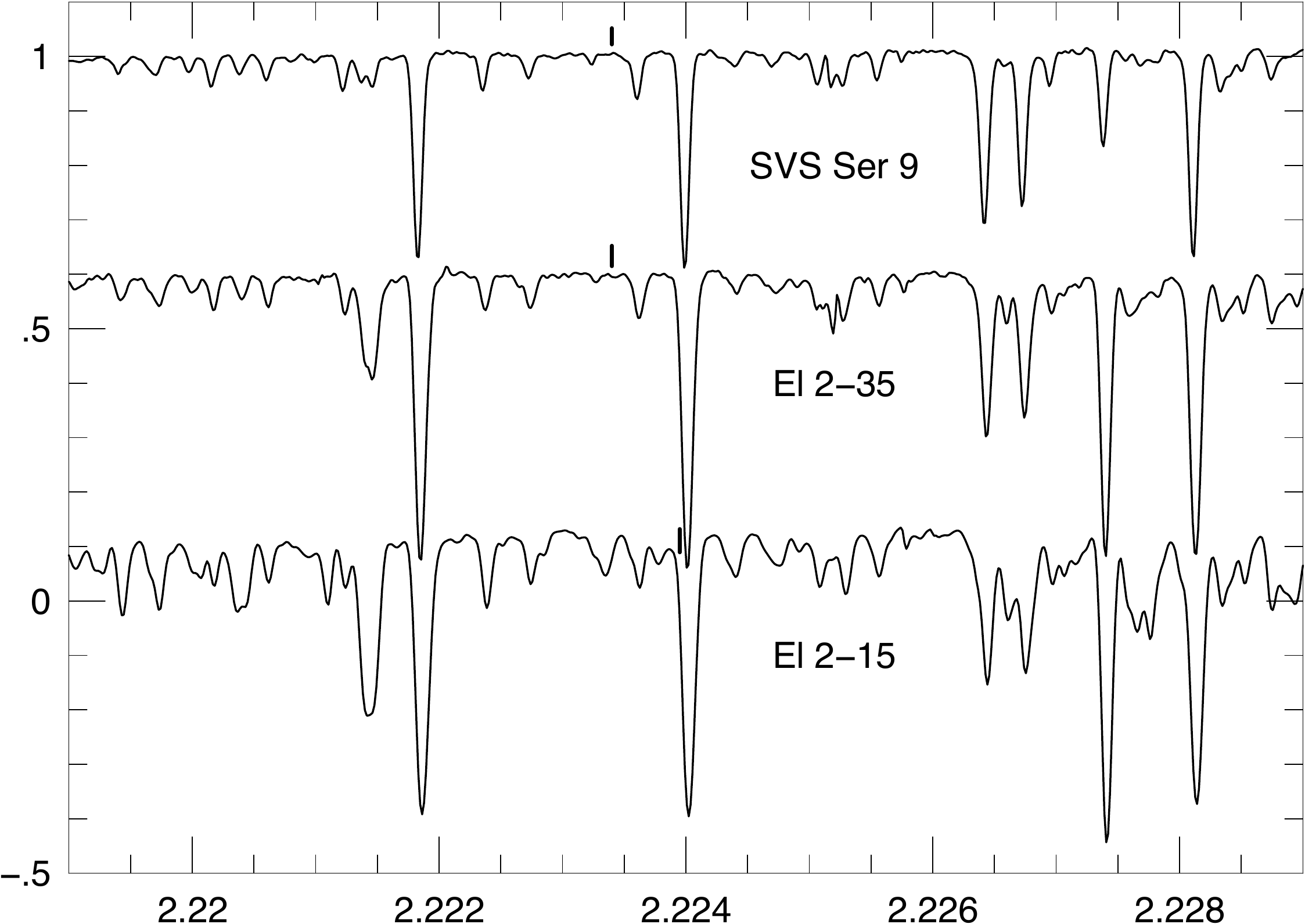}
\caption{Spectra of stars behind molecular clouds in Ophiuchus and Serpens in the \hh\ S(0) line region.
Positions of the S(0) line are marked with ticks (at the CO absorption velocities).
\label{H2Ophfig}}
\end{figure}

Spectra of the CO R-branch region toward the Taurus and Monoceros embedded sources are shown in Figure~\ref{COembdfig}.
Only the R(0)--R(3) lines and corresponding P-branch lines are seen toward Elias~3-1 and Elias~3-18.
The R(0)--R(8) lines are seen toward AFGL~989.
IRAS~04278+2253 shows a complicated spectrum, with narrow lines at R(0)--R(3) and two sets of broader lines at higher J.
It is difficult to identify the lines falling between the stronger high-J lines toward IRAS~04278+2253, with the low-J lines appearing to be missing.
These lines may be highly blueshifted, with the R(0) line possibly lying between the narrower R(3) and R(2) lines, near 2.34\um.
Broad absorption or emission lines may also be present toward Elias 3-1 and Elias 3-18.
Spectra of the \hh\ S(0) line region toward the Ophiuchus and Serpens sources are shown in Figure~\ref{H2Ophfig}, none of which shows clear evidence of \hh\ absorption.
Elias~2-15 and Elias~2-35 are cooler than the background stars in Taurus, resulting in complicated photospheric spectra.

\section{Spectrum Modeling}

\subsection{Smooth Spectrum Sources}

The hot star HD~283809 and the dust-enshrouded embedded sources Elias~3-1, Elias~3-18, IRAS~04278, and AFGL~989 have relatively flat, smooth spectra, simplifying their modeling.
The modeling procedure started with flat spectra on which the interstellar lines were superimposed.
The CO level populations were assumed to be described by a single-temperature Boltzmann distribution.
Because the lines are spectrally unresolved and may be somewhat optically thick, a curve of growth was used to calculate the line equivalent widths, and narrow lines with these equivalent widths were used in the models.
The line shapes in the curve of growth calculation were assumed to be Gaussian.
CO line strengths were taken from hitran \citep{hitran}.
Next, the source spectra with the interstellar lines superimposed were multiplied by a model for the telluric atmosphere transmission, and those spectra were convolved with the instrumental resolution.
Finally, the convolved source spectra were divided by a convolved model telluric spectrum to simulate the data reduction procedure in which the observed source spectra were divided by spectra of A0V comparison stars.
The procedure of first multiplying by a model telluric spectrum, convolving with the resolution function and then dividing by a convolved telluric spectrum, which might appear to be unnecessary, corrects for the fact that the telluric absorption that affects the interstellar line depths differs from the observed telluric absorption because the telluric lines are not fully resolved at our R=45,000 resolution.
The instrumental resolution and CO column densities, temperatures, and line widths were considered free parameters which were adjusted to give the best fit to the observed spectra.
The fit parameters are given in Table 2.

The effect of line saturation is very similar to that of non-LTE populations or a mixture of different temperatures along the line of sight if only R-branch lines are included in the modeling.
However, inclusion of the P-branch lines breaks this degeneracy as the optical depths of the R-branch lines, especially R(1), are greater than those of the corresponding P-branch lines.
Using line widths that fit the observed P-branch to R-branch equivalent width ratios, the CO spectra of HD~283809, Elias~3-1, and Elias~3-18 are well fitted by single temperature models, all with temperatures of $\sim$10 K.
The fitted line widths for these sources are in the range 0.4-0.5\,\kms.
The gas toward AFGL~989 is considerably warmer, T $\approx$ 35 K, and more turbulent;
the P/R line ratios are consistent with optically thin lines.
Some background sources, discussed below, also show broader lines.
Emission in the CO lines should be negligible, as gas temperatures are far too low to excite v=2 states, and radiatively excited states decay quickly via v=2-1 transitions.

\begin{figure}
\includegraphics[angle=270,scale=0.33]{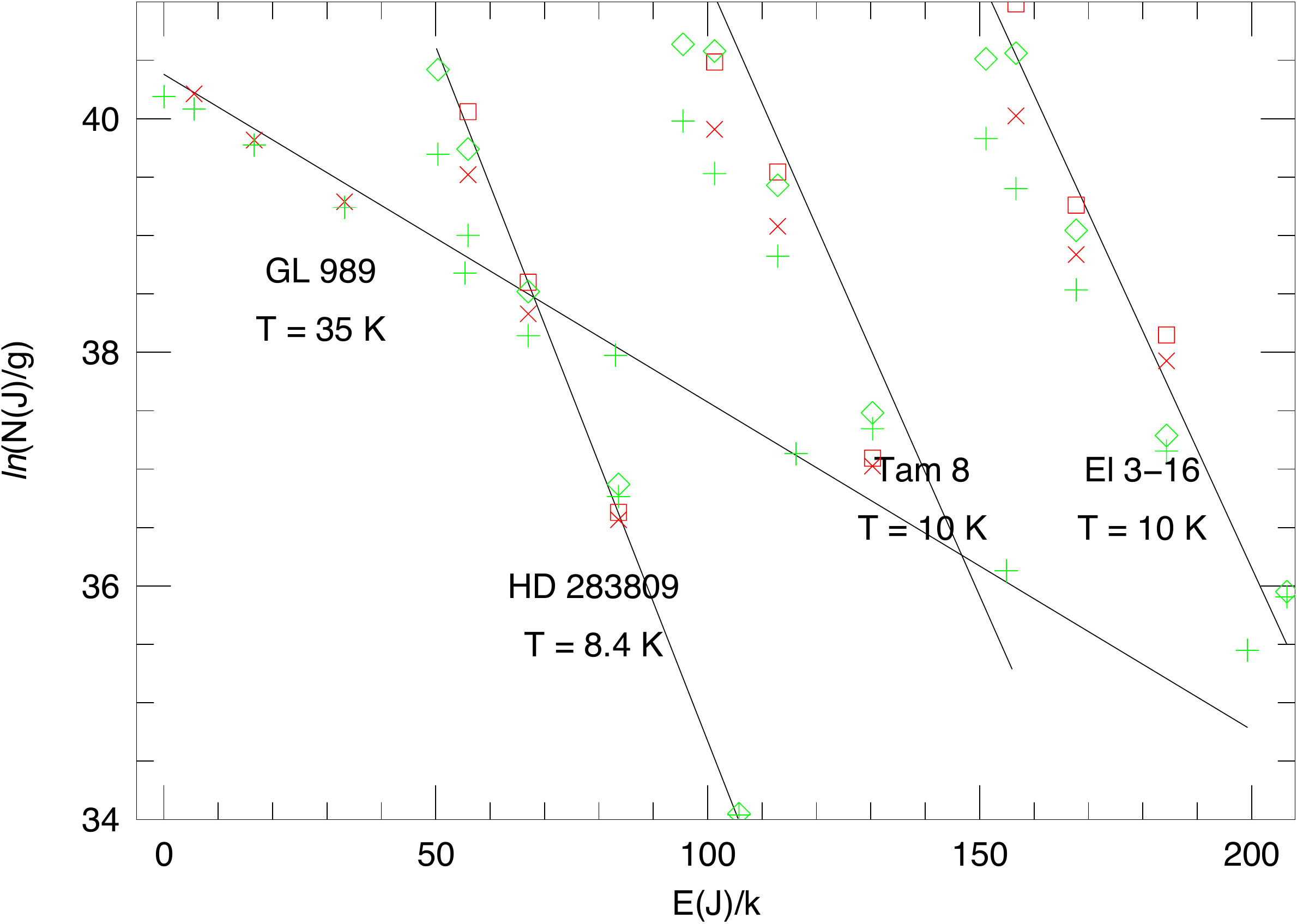}
\caption{Boltzmann diagram for CO toward AFGL~989, HD~283809, Tamura~8, and Elias~3-16.
Data for HD~283809, Tamura~8, and Elias~3-16 are offset horizontally by 50, 100, and 150 K, respectively.
X and + symbols are for P and R-branch lines, assuming the lines to be optically thin.
Diamonds and squares are for these lines after correction for saturation, assuming Gaussian line curves of growth, with line widths derived from the P(1)/R(1) and P(2)/R(2) equivalent width ratios.
Column densities are derived from measured equivalent widths.
Lines are from single temperature fits to the spectra as described in the text.
\label{NCOfig}}
\end{figure}

The column densities of the different rotational states of CO toward AFGL~989 and HD~283809 are shown in a Boltzmann diagram in Figure~\ref{NCOfig}.
For AFGL~989 the lines are broad enough that the correction for optical depth is small, and only the column densities assuming the lines to be optically thin are shown.
The derived column densities both with and without correction for saturation are shown for HD~283809.
The correction is substantial for the J=0 and J=1 lines, and significantly improves the fit to a single temperature population distribution.
The other two sources in this figure are discussed below.

A similar procedure was used to derive the \hh\ J=0 column density from the S(0) line, except that saturation of this weak line is negligible.
The line strength was calculated from \citet{turner77}.
The low temperature of the CO gas toward HD~283809, Elias~3-1, and Elias~3-18 indicates that only the J=0 state of \hh\ should be populated if the \hh\ ortho:para ratio is thermalized at the gas temperature.
Only the S(0) line was detected toward HD~283809, consistent with this expectation.
S(0) was not detected toward Elias~3-1 or Elias~3-18, but S(1) was detected in emission, and both lines were seen in emission from IRAS~04278, presumably from hot or fluorescent gas near these sources.
Because of the dominance of \hh\ emission toward these sources we did not attempt to derive the \hh\ column densities along these lines of sight.

\hh\ S(0) was seen in absorption toward AFGL~989, whereas S(1) and Q(1-4) were seen in emission.
The presence of both emission and absorption toward this source complicates its modeling, but the observation of Q-branch lines allows us to correct for emission in the S(0) line, albeit with substantial uncertainty.
We calculated the emission contribution to S(0) by multiplying the flux in the Q(2) emission line, which originates in the same v=1, J=2 upper state as S(0), by the branching ratio and a reddening factor.
Q(2) absorption should be negligible as the J=2 state should not be populated in the cold absorbing gas.
The reddening between Q(2) and S(0) was calculated based on the assumption that ${\rm A_V / N_{CO} = 5.5 \times 10^{-18}}$, the average of the ratio for the sources toward which we measured both.
We note that this corrected equivalent width is consistent with that observed by \citet{kulesa02} with a smaller aperture, which was probably less affected by emission.
We attempted to use the S(1) and Q(3) lines to measure the extinction to the source, assuming both are dominated by emission, but the large uncertainty in the Q(3) flux and the small wavelength difference between these lines resulted in too large an uncertainty to be useful.

\begin{deluxetable*}{ccccccccccc}
\tablecaption{Measured interstellar gas and dust absorption properties}
\tablewidth{0pt}
\tablehead{
\colhead{Name} & \colhead{N$_{\rm CO~thin}$} & \colhead{N$_{\rm CO~cog}$} &
\colhead{T$_{\rm CO}$} & \colhead{v$_{\rm LSR}$} &
\colhead{b$_{\rm CO}$\tablenotemark{a}} & \colhead{N$_{\rm H_2}$} & \colhead{J-K} &
\colhead{E$_{\rm J-K}$\tablenotemark{b}} & \colhead{A$_{\rm J}$\tablenotemark{c}}  &
\colhead{A$_{\rm V}$\tablenotemark{c}} \\
\colhead{} & \colhead{$10^{18}$ cm$^{-2}$} & \colhead{$10^{18}$ cm$^{-2}$} & \colhead{K} & \colhead{km~s$^{-1}$} & \colhead{km~s$^{-1}$} & \colhead{$10^{22}$ cm$^{-1}$} & \colhead{mag} & \colhead{mag} & \colhead{mag} & \colhead{mag}
}
\startdata
embedded stars &&&&&&&&&& \\
Elias~3-1 & 0.96 & 2.1 & 10.9 & 7.9 & 0.40 & undet & & & & 10.5\tablenotemark{d}\\
Elias~3-18 & 0.55 & 0.86 & 10.0 & 6.4 & 0.33: & undet & & & & 19.\tablenotemark{d} \\
IRAS 04278\tablenotemark{e} & 0.73\tablenotemark{e} & 1.0\tablenotemark{e} & 12.0\tablenotemark{e} & 9.3\tablenotemark{e} & 0.57\tablenotemark{e} & emiss & & & & \\
AFGL~989 & 4.1 & 4.4 & 35. & 50.5 & 3.1: & 1.4\tablenotemark{f} & & & & \\
background stars &&&&&&&&&& \\
Elias~3-13 & 0.90 & 1.9 & 9.1 & 6.4 & 0.32 & 1.5 & 3.05 & 2.30 & 3.7 & 12.4 \\
Elias~3-16 & 1.4 & 2.6 & 9.8 & 6.8 & 0.49 & 1.5 & 5.45 & 4.50 & 7.2 & 24.3 \\
Tamura~8 & 1.3 & 2.5 & 9.5 & 6.4 & 0.61 & 1.7 & 5.07 & 4.49 & 7.2 & 24.2 \\
HD~283809 & 0.84 & 1.3 & 8.4 & 6.0 & 0.41 & 0.75 & 0.83 & 0.83 & 1.33 & 4.5 \\
Kim~1-59 & 0.8 & 1.1 & 10.0 & 6.4 & 1.5 & 0.9: & 2.87 & 2.12 & 3.4 & 11.4 \\
SVS76 Ser 9 & 2.7 & 2.9 & 12.9 & 0.0 & 1.1 & 0.9: & 4.33 & 3.48 & 5.6 & 18.8 \\
\enddata
\tablenotetext{a}{Doppler b parameter = $e^{-1}$ HW.  FWHM = 1.66 b}
\tablenotetext{b}{Intrinsic J-K colors of K giants based on fitted \teff.
Intrinsic colors of HD~283809 assumed = 0.0 mag.}
\tablenotetext{c}{Using R$_V$ = 5.5 extinction curve of \citet{WD01}, $\rm A_J / E_{J-K} = 1.60$,
and $\rm A_V / E_{J-K} = 5.4$.}
\tablenotetext{d}{\citet{TE99}}
\tablenotetext{e}{Multiple velocity and temperature components are seen toward IRAS 04278+2253.
A fit to the J = 0-3 CO lines, which are dominated by the coldest component, is given here.}
\tablenotetext{f}{Based on S(0) equivalent width corrected for emission based on Q(2) emission.}
\end{deluxetable*}

\subsection{Late-Type Stellar Spectrum Modelling}

Except for HD~283809, the background stars (Elias~3-13, Elias~3-16, Kim~1-59, Tamura~8, Elias~2-15, Elias~2-35, and SVS Ser~9) are all late-type giants, which have complicated photospheric spectra.
Consequently, it was necessary to remove photospheric features from the observed spectra to measure the interstellar lines.

To generate synthetic spectra to model the background stars, we used the current version
of the LTE line analysis code MOOG \citep{sneden73}\footnote{
Available at http://www.as.utexas.edu/$\sim$chris/moog.html}.
Line lists for these calculations were generated from current molecular
laboratory data (CN, \citealt{sneden14}; OH \citealt{brooke15}; CO,
\citealt{goorvitch94}, with bandstrengths empirically increased by
0.15~dex as discussed in \cite{afsar16}) and current atomic laboratory
data (\eg, \citet{sneden16}, and references therein).
However, atomic lab data are not plentiful in the H and K bands, so we
supplemented these transitions with those of the \cite{kurucz11} line
compendium\footnote{
Available at http://kurucz.harvard.edu/linelists.html}.
We interpolated stellar model atmospheres from the ATLAS grid
\citep{kurucz11}\footnote{
Available at http://kurucz.harvard.edu/grids.html; model interpolation
software kindly provided by Andy McWilliam and Inese Ivans.},
using model parameters to be discussed below.
The computed spectra were smoothed empirically with Gaussians to match the
observed line profiles.
The smoothing FWHM were typically $\simeq$0.065~nm; this includes contributions
from the spectrograph slit function ($\sim$0.050~nm) and stellar
macroturbulence ($\sim$0.040~nm) added in quadrature.

\subsection{Analyses of Foreground Red Giants\label{localstars}}

The five foreground red giants serve two purposes: as standards for our stellar
computations and for determination of residuals to the model spectra,
which we use to correct the models of the background stars.
Their HIPPARCOS-based parallaxes \citep{vanleeuwen07} and spectral types
were taken from the SIMBAD database \citep{wenger00};
these are listed in Table~\ref{stdparam}.
The mean distance of these stars is $\simeq$165~pc, and thus they are close
enough to be essentially un-reddened.
Stellar parameters for the foreground red giants have been reported
in the recent literature.
All have \teff\ estimates in the very large-sample spectral energy
distribution study of \cite{mcdonald12}.
Four of the five stars are included in the PASTEL stellar parameter
catalog \citep{soubiran10,soubiran16}.
Their \teff, \logg, and [Fe/H] values\footnote{
We adopt the standard stellar spectroscopic notation \citep{wallerstein59}
that for elements A and B,
[A/B] $\equiv$ log$_{\rm 10}$(N$_{\rm A}$/N$_{\rm B}$)$_{\star}$ $-$
log$_{\rm 10}$(N$_{\rm A}$/N$_{\rm B}$)$_{\odot}$.
We use the definition
log~$\epsilon$(A) $\equiv$ log$_{\rm 10}$(N$_{\rm A}$/N$_{\rm H}$) + 12.0,
and equate metallicity with the stellar [Fe/H] value.}
are adopted here and listed in Table~\ref{stdparam}.
We use these parameters as fundamental anchors for all of the program stars.

HR~7800 is not in the PASTEL catalog.
\cite{mcdonald12} derived \teff~=~3832~K for this star.
For our stars with \teff\ estimates by both PASTEL and \citeauthor{mcdonald12},
the mean difference is
$\langle \teff_{PASTEL}-\teff_{McDonald} \rangle$ $\simeq$ $-$175~K.
We therefore adjusted the HR~7800 temperature to \teff~=~3760~K.
We lack independent knowledge of this star's other atmospheric parameters.
We extrapolated downward in temperature from the other foreground giants, and
adopted a rough gravity estimate of \logg~= 1.40.
In Table~\ref{stdparam} we see that all stars with PASTEL parameters are
slightly metal-poor, $\langle$[Fe/H]$\rangle$ $\simeq$ $-$0.15.
We adopted this value also for HR~7800.
Finally, we adopted a uniform microturbulent velocity of 2.5~\kms\
for all of these cool red giants.

The CNO group elements are coupled via molecule formation, and their abundances
ideally should be derived iteratively.
Our goal for both the foreground and the background stars is simpler,
to match synthetic and observed stellar spectra to
effect a cancellation of these spectra.
Therefore we first simplified the process by assuming [O/Fe] for each star.
This is justified for metal-rich red giants; see many surveys,
\eg, \cite{lambert81}, \cite{kjaergaard82}, \cite{afsar12}, \cite{holtzman15}.
Then we synthesized the CO-dominated wavelength region 2.326$-$2.355~\um ,
where the CO ISM lines are also located.
Over most of this region the main stellar features are individual lines of
the $^{12}$CO v=3-1 band, with additional contributions from 2-0 lines.
The R-branch bandhead of the $^{13}$CO 2-0 band begins at about
2.334~\um , and the $^{12}$CO 4-2 bandhead begins at about 2.353~\um.
Carbon isotopic ratios vary widely over the range
5~$\lesssim$ $^{12}$C/$^{13}$C $\lesssim$~30 in high metallicity red giants.
In principle we could derive carbon isotopic ratios for all of the stars.
However, almost all $^{12}$CO lines are very strong (saturated) in our
stars while the $^{13}$CO features are much weaker.
Therefore there is an interplay between adopted microturbulent velocities
and derived $^{12}$C/$^{13}$C values.
Proper resolution of this issue would involve synthesizing a much larger
wavelength range, and careful attention to excitation potential strength
dependences of the $^{12}$CO lines.
This exercise is beyond the scope of our work.
We do see significant $^{13}$CO absorption in all stars, and after some
numerical experiments we adopted uniformly $^{12}$C/$^{13}$C~=~15 for
our syntheses.

We varied the carbon abundances until best synthetic/observed matches
were achieved.
After determining the carbon we then used similar observed
spectrum matching to derive stellar nitrogen abundances from CN lines located
in the 2.21-2.24~\um\ region, which also includes the ISM H$_2$ v=1-0 S(0).

In Table~\ref{stdparam} we list the derived [C/Fe] and [N/Fe] abundances
for the five foreground red giants, adopting the \cite{asplund09} solar abundances
for computations of these abundance ratios.
We have grouped the four stars with PASTEL parameters and have listed
their mean values before presenting the quantities separately for HR~7800.
[O/Fe] is also entered in the table to emphasize its defined value of 0.00.
It is apparent that the PASTEL red giants have nearly identical abundances,
with $\langle$[C/Fe]$\rangle$~= $-$0.34 and
$\langle$[N/Fe]$\rangle$~= $+$0.61.
with small star-to-star scatter.
HR~7800, with parameters based on shifting its literature \teff\ to be
consistent with the PASTEL stars, has [C/Fe] and [N/Fe] values in good
agreement with the means of the other four stars within mutual uncertainties.
These abundances are also in accord with previous studies that
have found deficiencies of carbon and overabundances of nitrogen in high
metallicity thin disk red giants, \eg, \cite{lambert81}, \cite{kjaergaard82},
\cite{taut10}, \cite{afsar12}.

The residuals of our fits to the foreground star spectra are highly correlated.
We attribute this to systematic errors in the models, due to missing lines and
erroneous line strengths or positions, and we expect the same errors to occur in
our models for the background stars.
Consequently, we averaged the residuals of the foreground star models
and added a multiple of this average residual spectrum to the background
star models to minimize their residuals, as is discussed below.

\begin{center}
\begin{deluxetable*}{lcrrrrrrrrr}
\tablewidth{0pt}
\tablecaption{Parameters of Foreground Red Giants\label{stdparam}}
\tablecolumns{11}
\tablehead{
\colhead{Name}                     &
\colhead{d}                        &
\colhead{Sp Type}                  &
\colhead{\teff\tablenotemark{a}}   &
\colhead{\teff\tablenotemark{b}}   &
\colhead{\logg\tablenotemark{b}}   &
\colhead{[Fe/H]\tablenotemark{b}}  &
\colhead{\vmicro}                  &
\colhead{[C/Fe]\tablenotemark{c}}  &
\colhead{[N/Fe]\tablenotemark{c}}  &
\colhead{[O/Fe]\tablenotemark{c}}  \\
\colhead{}                         &
\colhead{pc}                       &
\colhead{}                         &
\colhead{K}                        &
\colhead{K}                        &
\colhead{cm s$^{-2}$}              &
\colhead{}                         &
\colhead{km s$^{-1}$}              &
\colhead{}                         &
\colhead{}                         &
\colhead{}                   
}
\startdata
\mbox{HR 5899} &  115    &  K5 III & 4100 & 3920 & 1.68 & $-$0.17 &    2.5 & $-$0.30 & $+$0.65 & 0.00 \\
\mbox{HR 7919} &  138    &  K2 III & 4557 & 4485 & 2.40 & $-$0.08 &    2.5 & $-$0.45 & $+$0.55 & 0.00 \\
\mbox{HR 7956} &  136    &  K3 III & 4388 & 4190 & 2.12 & $-$0.12 &    2.5 & $-$0.30 & $+$0.60 & 0.00 \\
\mbox{HR 7969} &  180    &  K5 III & 4256 & 4010 & 1.78 & $-$0.23 &    2.5 & $-$0.30 & $+$0.65 & 0.00 \\
\mbox{mean}    & \nodata & \nodata & 4325 & 4151 & 2.00 & $-$0.15 &    2.5 & $-$0.34 & $+$0.61 & 0.00 \\
$\sigma$       & \nodata & \nodata &  194 &  249 & 0.33 &    0.06 & \nodata &    0.07 &    0.05 & \nodata \\
               &         &         &      &      &      &         &      &         &         &      \\
\mbox{HR 7800} &  259 &  K7 III & 3932 & 3760 & 1.40 & $-$0.15 &      & $-$0.30 & $+$0.50 &      \\
\enddata                                                      

\tablenotetext{a}{\cite{mcdonald12}}
\tablenotetext{b}{\cite{soubiran16}}
\tablenotetext{c}{\rm log ratio to solar abundances from \cite{asplund09}: 
log~$\epsilon$(C)~= 8.43,
log~$\epsilon$(N)~= 7.83,
log~$\epsilon$(O)~= 8.69.}                                      

\end{deluxetable*}                                             
\end{center}

\begin{center}
\begin{deluxetable*}{lcrrrrrrr}
\tablewidth{0pt}
\tablecaption{Parameters of Background Stars\label{progparam}}
\tablecolumns{9}
\tablehead{
\colhead{Name}                     &
\colhead{Sp Type\tablenotemark{a}} &
\colhead{\teff}                    &
\colhead{\logg}                    &
\colhead{[Fe/H]}                   &
\colhead{\vmicro}                  &
\colhead{[C/Fe]\tablenotemark{b}}  &
\colhead{[N/Fe]\tablenotemark{b}}  &
\colhead{[O/Fe]\tablenotemark{b}}  \\
\colhead{}                         &
\colhead{}                         &
\colhead{K}                        &
\colhead{cm s$^{-2}$}              &
\colhead{}                         &
\colhead{km s$^{-1}$}              &
\colhead{}                         &
\colhead{}                         &
\colhead{}
}
\startdata
\mbox{Elias 3-13}   & K2 III   & 4400 & 2.20 & $-$0.15 & 2.5 & $-$0.35 & $+$0.55 & 0.00 \\
\mbox{Elias 3-16}   & K5 III   & 3950 & 1.70 & $-$0.15 & 2.5 & $-$0.30 & $+$0.60 & 0.00 \\
\mbox{Tamura 8}   & G7 III   & 4900 & 2.80 & $-$0.15 & 2.0 & $-$0.30 & $+$0.70 & 0.00 \\
\mbox{Kim 1-59}    & K2 III   & 4400 & 2.20 & $-$0.15 & 2.5 & $-$0.25 & $+$0.30 & 0.00 \\
\mbox{SVS76 Ser 9} & K4III & 4150 & 2.00 & $-$0.15 & 2.5 & $-$0.30 & $+$0.70 & 0.00 \\
\mbox{Elias 2-35}   & K6 III   & 3800 & 0.50 & $-$0.15 & 2.5 & $-$0.25 & $+$0.40 & 0.00 \\
\mbox{mean}        &          &      &      &         &     & $-$0.29 & $+$0.54 &      \\
\enddata

\tablenotetext{a}{based on fitted \teff\ values.}
\tablenotetext{b}{ log ratio to solar abundances from \cite{asplund09}:
log~$\epsilon$(C)~= 8.43,
log~$\epsilon$(N)~= 7.83,
log~$\epsilon$(O)~= 8.69.}

\end{deluxetable*}
\end{center}

\subsection{Analyses of Background Red Giants\label{programstars}}

The background stars lie in heavily extincted sight lines.
Therefore they are faint in the optical spectral region; thus they lack
Hipparcos parallaxes and have not been treated to high-resolutions
spectroscopic analysis.
All of the background stars lie close to the Galactic plane
($b$~$\leq$~16$^\circ$).
Therefore they are almost surely the same kinds of thin-disk giants that
make up our foreground sample.
We made this assumption, and extended it to assert that, for the purposes of
this ISM study, the regularities of the the foreground red giants apply to
the background stars.

In particular, we assumed that the background stars have [Fe/H]~= $-$0.15, and
$^{12}$C/$^{13}$C~=~15.
we again used microturbulent velocities of 2.5~\kmsec,
except for the warmer star Tamura~8, for which we took 2.0~\kmsec\
as a reasonable estimate.
Further, we desired to end up with approximately the same [C/Fe] and [N/Fe]
values as the means in the foreground sample.
CO formation is very sensitive to temperature due to its large dissociation
energy (D$_0$~= 11.1~eV), and to a lesser extent so is CN (D$_0$~= 7.8~eV).
Therefore we computed trial synthetic/observed spectrum matches,
altering the \teff\ (with \logg\ changes to match) until [C/Fe]~$\simeq$ $-$0.3
was achieved.
Then these models were applied to the CN spectra, and the best-fit [N/Fe]
was adopted.
The results of these exercises are listed in Table~\ref{progparam}.
Although the  carbon abundances were forced to agree with the mean value of
the foreground sample, the nitrogen abundances were assessed independently
(within the constraints of the steps leading to derivation of the carbon),
and their mean is in reasonable agreement with that of the foreground stars.

\begin{figure}
\plotone{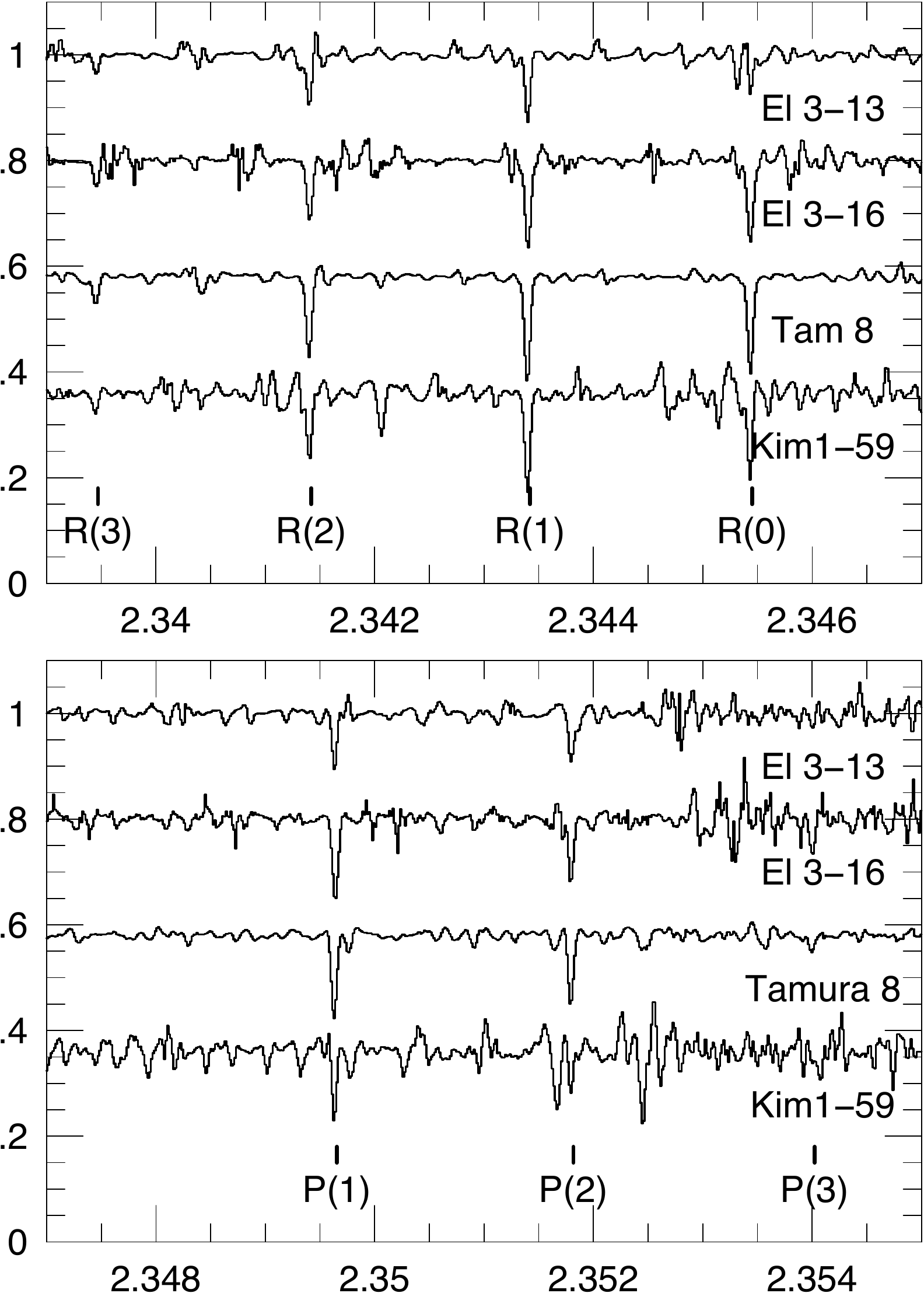}
\caption{Spectra of late-type stars behind the Taurus Molecular Cloud in the CO v=2-0 band-center region after correction for photospheric and telluric spectral features.
Positions of the interstellar CO lines are marked with ticks at v$_{LSR}$ = +7 \kms.
Spectra are normalized to a continuum level of 1.0 and offset.
\label{COfixdfig}}
\end{figure}

\begin{figure}
\plotone{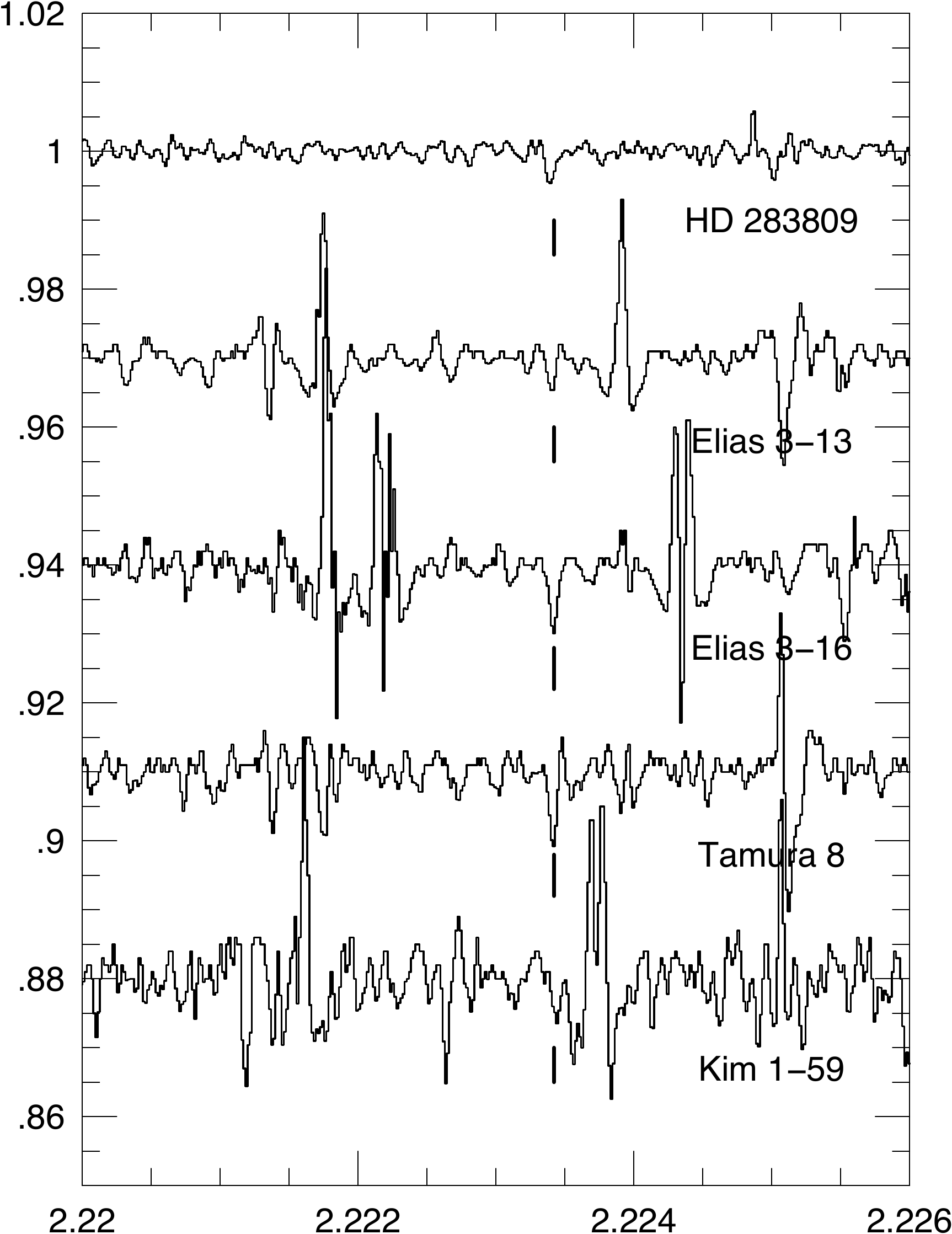}
\caption{Spectra of stars behind the Taurus Molecular Cloud in the \hh\ v=1-0 S(0) line region after correction for photospheric and telluric spectral features.
Positions of the S(0) line are marked with ticks at the CO band \vlsr\ of +7 \kms.
Although some residuals are greater than the interstellar lines, almost all are at wavelengths of strong photospheric features or telluric absorption.
\label{H2fixdfig}}
\end{figure} 

\subsection{Background star and ISM parameters}

To derive interstellar gas properties the observed background star spectra were first modeled with a procedure like that used for the flat-spectrum sources, but starting with the best-fitting model photospheric spectra.
The spectra of the Taurus background sources, after division by the model photospheric spectra, including a fraction of the foreground star residual spectrum, are shown in Figures \ref{COfixdfig} and \ref{H2fixdfig}.
Rotational level populations derived from the spectra of Tamura~8 and Elias~3-16 after the modeling described above, and both before and after correction for optical depths of the CO lines, are shown in Figure~\ref{NCOfig}.
 
Since at most four CO rotational states (J=0-3) had detectable absorption lines toward these stars, two-temperature fits could not be justified, and the single temperature models provided good fits to the observations.
CO absorption was detected toward all sources in Table 1, but the quality of the fits to the photospheric spectra of Elias 2-15 and Elias 2-35 were too poor to allow measurement of the CO parameters, probably as a result of the low temperatures of these stars.
\hh\ S(0) absorption was reliably detected toward Elias~3-16 and Tamura~8, probably detected toward Elias 3-13 and HD~283809, and possibly present toward Kim~1-59.
None of the background sources showed evidence of \hh\ S(1) absorption (or emission), as is expected given the low gas temperatures ($\sim$10 K) derived from the CO spectra.

\section{Interstellar gas properties}

Derived interstellar absorption parameters are given in Table 2.
In addition to the fit parameters, estimates of the dust extinction to the sources are given.
Extinctions for the background K giant stars are based on colors from 2MASS \citep{cutri03}
with intrinsic colors based on our fitted effective temperatures with the \teff\ -- color relationship of \citet{alonso99}.
The intrinsic J-K color of the hot star HD~283809 is assumed to be zero.
SIMBAD quotes a spectral type of F8.
This spectral type corresponds to (J$-$K)~$\simeq$ 0.3 \citep{alonso99}.
However, \citet{straiz82} used medium-resolution spectroscopy
to demonstrate that HD~283809 is a B-type star, for which (J$-$K)~$\simeq$ 0.
The \citet{WD01} $\rm R_V$ = 5.5 extinction curve, which gives $\rm A_V / E_{J-K} = 5.4$, was used to calculate \av.

No uncertainties are given for the quantities in Table 2, as the statistical uncertainties in the fits are generally small compared to systematic uncertainties, which are difficult to quantify.
The CO column densities, ${\rm N_{CO\,cog}}$, depend strongly on the value of the CO line width,
\bco , through its effect on the curve of growth.
The need for a curve of growth correction to the column densities derived assuming the lines to be optically thin can be seen in Figure~\ref{COfixdfig}.
If the CO lines were optically thin, the R(1) line would be twice as strong as the P(1) line, whereas the observed ratio is closer to unity.
The value of \bco\ is constrained primarily by this ratio (and to a lesser extent by the observed R(2)/P(2) ratio, which in optically thin gas would be 1.5).
Unfortunately, these lines are in a spectral region with substantial photospheric (and telluric) absorption, so the derived \bco\ value depends on the quality of the stellar models.
The similar values of \bco\ for the high S/N Taurus cloud sources (including HD~283809, which did not require stellar modeling) encourage us that these values are reasonably reliable, but we would guess that they could easily be in error by $\pm 0.1$ \kms , resulting in errors in \nco\ $\sim$20\%.
Errors in ${\rm T_{CO}}$ due to the uncertainty in \bco\ are typically $\pm$2 K.

Statistical uncertainties and stellar modeling uncertainties make comparable contributions to the possible errors in \nhh\ for the background K giants.
We estimate that \nhh\ for Elias~3-16 and Tamura~8 is uncertain by $\sim$25\%.
\nhh is uncertain by $\sim$50\% for Elias~3-13 and HD~283809.
We consider \hh\ to be only marginally detected for Kim~1-59 and SVS Ser 9.
The \hh\ S(0) line is clearly detected toward GL 989 (and was previously detected by \citet{kulesa02}), but the correction for S(0) emission adds considerable uncertainty to our \nhh\ determination.
We estimate the uncertainty in \nhh\ to be $\pm$50\% (based on our measurements; the fact that \citet{kulesa02} measured a very similar number with a smaller beam that would be less affected by emission indicates that the actual error is probably smaller than this).

The primary sources of uncertainty in the extinction values for the background stars are the intrinsic colors of the K giants and the assumption of the \citet{WD01} $\rm R_V$ = 5.5 extinction curve.
We use the colors of \cite{alonso99} with our fitted \teff\ values to determine the intrinsic colors.
The \citet{WD01} $\rm R_V$ = 5.5 extinction curve is fitted to molecular cloud observations, so should be appropriate for our lines of sight.
Their R$_V$ = 3.1 curve gives an only slightly larger $\rm A_V / E_{J-K}$ of 5.5.

\begin{deluxetable*}{cccccccccc}
\tablecaption{Column density and extinction ratios}
\tablewidth{0pt}
\tablehead{
\colhead{Name} & \colhead{$\rm N_{COgas}$} & \colhead{$\rm N_{COsolid}$} & \colhead{\nhh }  & \colhead{\av } & \colhead{COgas/\av } & \colhead{COtot/\av } & \colhead{\hh /\av } &
\colhead{\hh /COgas} & \colhead{\hh /COtot} \\
\colhead{} & \colhead{$\rm 10^{18} cm^{-2}$} & \colhead{$\rm 10^{18} cm^{-2}$} & \colhead{$\rm 10^{22} cm^{-2}$} & \colhead{mag} & \colhead{$\rm 10^{17} cm^{-2}$} & \colhead{$\rm 10^{17} cm^{-2}$} & \colhead{$\rm 10^{21} cm^{-1}$} & \colhead{} & \colhead{}
}
\startdata
Elias~3-13 & 1.9 & 0.11\tablenotemark{a} & 1.5 & 12.4\tablenotemark{d} & 1.53 & 1.6 & 1.2 & 7900 & 7500 \\
Elias~3-16 & 2.6 & 0.74\tablenotemark{a} & 1.5 & 24.3\tablenotemark{d} & 1.07 & 1.4 & 0.67 & 5800 & 4500 \\
Tamura~8 & 2.5 & 0.65\tablenotemark{a} & 1.7 & 24.2\tablenotemark{d} & 1.03 & 1.3 & 0.70 & 6800 & 5400 \\
HD~283809 & 1.3 & 0.\tablenotemark{b} & 0.75 & 4.5\tablenotemark{e} & 2.9 & 2.9 & 1.7 & 5800 & 5800 \\
Kim~1-59 & 1.1 & 0.4\tablenotemark{b} & 1.6\tablenotemark{c} & 10.3\tablenotemark{d} & 1.07 & 1.5 & 0.77\tablenotemark{c} & 14,500\tablenotemark{c} & 10,700\tablenotemark{c} \\
AFGL~989 & 4.0 & - & 1.4 & - & - & - & - & 3500 & 3500 \\
SVS Ser 9 & 2.9 & 0.8\tablenotemark{b} & 0.9\tablenotemark{c} & 18.8\tablenotemark{d} & 1.54 & 2.0 & 0.48\tablenotemark{c} & 3100\tablenotemark{c} & 2400\tablenotemark{c} \\
Average\tablenotemark{f} & & & & & 1.52 & 1.8 & 1.07 & 6600 & 5800 \\
\enddata
\tablenotetext{a}{\citet{TE99}.  \citet{chiar95} give slightly lower numbers, but \citet{whittet07} point out that solid CO$_2$ should be included as CO on grains may be converted to CO$_2$.}
\tablenotetext{b}{Estimated from \av\ based on formula from \citet{whittet07}.}
\tablenotetext{c}{Uncertain.  Consistent with 0.}
\tablenotetext{d}{Extinction derived from J-K colors, based on fitted \teff\ and $\rm A_V / E_{J-K}$ = 5.4.}
\tablenotetext{e}{Intrinsic J-K assumed = 0.}
\tablenotetext{f}{Average of more reliably determined ratios.  Anomolous value of gas/dust toward HD 283809 increases the ratios to \av.}
\end{deluxetable*}

\section{Discussion and Conclusions}

The measured column densities of \hh\ and CO, the extinctions, and the ratios of these quantities are given in Table 5. 
All three quantities were measured toward Elias~3-13, Elias~3-16, Tamura~8 and HD~283809.
\nhh\ and \nco\ were measured toward AFGL~989, but the extinction is unknown.
\nco\ and \av\ were measured toward Kim~1-59, and SVS~Ser~9, but \nhh\ is quite uncertain.

\subsection{\nhh , \nco , and \av }

For the four sources where \nco\ and \av\ were most reliably measured, all of which lie behind the Taurus Molecular Cloud, N$_{\rm COgas}$/\av\ = 1.0 -- $2.9 \times 10^{17}$ cm$^{-2}$, with an average of $1.5 \times 10^{17}$ cm$^{-2}$ and N$_{\rm COtot}$/\av\ = 1.3 -- $2.9 \times 10^{17}$ cm$^{-2}$, with an average of $1.8 \times 10^{17}$ cm$^{-2}$, where $\rm N_{COtot}$ includes solid CO.
\citet{FLW82} measured $\rm N_{C^{18}O} / A_V = 1.7 \times 10^{14} cm^{-2}$, which corresponds to $\rm N_{COgas} / A_V = 0.85 \times 10^{17} cm^{-2}$, lower than any of our values and a factor of $\sim$1.8 below our average.
\citet{pineda10} measured $\rm N_{COgas} / A_V = 1.0 \times 10^{17} cm^{-2}$, a factor of 1.5 below our average, and $\rm N_{COtot} / A_V = 1.4 \times 10^{17} cm^{-2}$, a factor of 1.3 below our average.

For the four sources where \nhh\ and \av\ were most reliably measured, \hhav\ = 0.7 -- $1.7 \times 10^{21}$ cm$^{-2}$ and the average is $1.07 \times 10^{21}$ cm$^{-2}$.
These sources lie behind the Taurus Molecular Cloud (TMC).
Including the sources with more uncertain values of \nhh\ would increase the range and lower the average to 0.92.
Our average value of \hhav\ is consistent with the assumption that the diffuse interstellar medium value of ${\rm N_H / A_V}$ of $1.9 \times 10^{21}$ cm$^{-2}$ \citep{BSD78} is valid in molecular clouds and that essentially all H atoms reside in \hh\ molecules.

For the four Taurus background sources where \nco\ and \nhh\ were most reliably measured, $\rm N_{H_2} / N_{COgas}$ ranges from 5800 to 7900, with an average of 6600, and $\rm N_{H_2} / N_{COtot}$ ranges from 4500 to 7500, with an average of 5800.
AFGL~989, in Monoceros, has a lower \hhco\ value, but because it is an embedded star, which required a correction for \hh\ emission, we do not conclude that the difference necessarily indicates a difference between the two clouds.
Kim~1-59 (behind the TMC) appears to have a higher ratio, and SVS~Ser~9 (in Serpens) appears to have a lower ratio, but the detection of \hh\ absorption in these sources is quite uncertain.
Our \hhco\ value is consistent with the value measured toward high-mass embedded young stars by \citet{lacy94} , \citet{kulesa02} , and \citet{goto15}.
As was concluded by those authors, the \hhco\ ratio is a factor $\sim$2 lower than derived from mm-wave CO emission and star counts \citep{FLW82}.
\citet{goto15} suggest that there is a trend in this ratio, with \hhco\ increasing with Galactocentric distance.
All of our sources lie within 1 kpc of the Sun, so we cannot test this conclusion.
Our average measured value of \hhco\ (5800) corresponds to $\sim$30\% of the total interstellar carbon being in CO, assuming the solar C/H ratio, and 60\% of gas-phase carbon being in CO, assuming 50\% of carbon resides in grains.
This is more than has been assumed in the past, but it still leaves some of the interstellar carbon unaccounted for.

\subsection{Variations in ratios and uncertainties}

Are the variations in \coav , \hhav , and \hhco , either among the TMC sources or between different molecular clouds, significant?
The values of \coav\ and \hhav\ toward HD~283809 appear to be significantly higher than toward the other sources, although the \hhco\ ratio is in the middle of the range observed.
This difference is probably real.
It could result from an underestimate of $\rm A_V$, but this would require a negative intrinsic J-K color, which seems unlikely.
Among the Taurus sources other than HD~283809 variations in the ratios are probably not significant.
The observations of sources toward Monoceros, Ophiuchus, and Serpens are not sufficiently reliable to make a strong statement about any variations between molecular clouds.

Our conclusion that \hhco\ is lower than generally assumed is consistent with the conclusion based on observations of high-mass embedded stars.
But our conclusion that \coav\ has been underestimated, rather than \hhav\ being overestimated is more surprising.
The largest uncertainty in our determination of \nco\ is in the curve of growth correction for optical depth in the CO lines.
This correction is based on the ratios of equivalent widths of P and R-branch lines, especially P(1) and R(1), and the assumption of Gaussian line shapes.
However, it is apparent from inspection of the data that a saturation correction is required, and a line shape with stronger wings, such as a Lorentzian, would require a greater CO column density, and so an even lower \hhco\ ratio.
The determination of \av\ depends on the intrinsic colors of the background stars and the assumed extinction law.
However, J-K colors of K2-K5 III stars vary by only $\pm$0.12 mag, and the value of $\rm A_V / E_{J-K}$ does not vary significantly with $\rm R_V$.
For the \coav\ ratios measured by \citet{dickman78}, \citet{FLW82}, and \citet{pineda10} to be erroneously low would require that rotational line emission underestimates \nco\ or that star counts overestimate \av.
\citet{pineda10} find variations in \coav\ with location in the TMC.
To test whether our observations were made at atypical locations we compared our values of \nco\ and \av\ to theirs in the same directions (although at spatial resolution of 40\as\ for \nco\ and 200\as\ for \av ), using data provided by Pineda (private communication).
The values of \av\ toward our stars were 0.8-1.8 times theirs, a difference consistent with likely spatial variations.
However, our values of \nco\ were consistently larger than theirs, with a ratio of 1.5-4.5, and our average \coav\ ratio was 2.1 times theirs.
Since our measurements of \nco\ and \av\ were necessarily along the same lines of sight we don't think observational bias or selection effects should have affected our value of \coav.

\subsection{Comparison with chemical models}

There are many studies of molecular abundances in interstellar clouds.
Near the cloud surfaces, in PDRs, the models predict that hydrogen is predominantly molecular beyond $\rm A_V \sim 1$ and that carbon becomes first neutral and then molecular (predominantly CO) at $\rm A_V \sim 3$ \citep{HT99}.
If all H is in \hh\ and all C is in CO, \hhav $\rm \sim 10^{21} cm^{-2}$ and \hhco\ $\sim$ 3500, assuming 50\% of carbon is in grains.
Beyond $\rm A_V \sim 10$ an increasingly large fraction of CO is observed to freeze out onto grains \citep{whittet07}.
But this simple layered structure is somewhat contradicted by the observation of [C~I] emission from deep within molecular clouds \citep{plume94}.
\citet{glover10} have made a model of \hh\ and CO abundances in turbulent molecular clouds, in which the non-uniform density structure allows UV photons to penetrate into the clouds where they can dissociate CO.
They do not include CO freeze-out, but their model should be a better approximation to molecular cloud structure than uniform density models, and since the observed solid CO abundance along our lines of sight are small compared to the gas-phase CO the neglect of freeze-out should not be a serious problem for our sources.
They find highly spatially variable column densities and \hhco\ ratios in their models,
but their mass-weighted mean \hhco\ ratios are similar to our observed ratios, and their spatially resolved \hhco\ ratios are similar to ours where their \hh\ column densities are similar to ours.
They assume a much larger turbulent velocity than we observe through the TMC (5\,\kms\ vs. $\sim$0.5 \kms\ rms), but otherwise their model appears to fit our observations reasonably well.
Their conclusion that \hhco\ and \hhav\ vary across a cloud tends to support the reality of the
variations we observe, although we might expect smaller variations in the more quiescent TMC.
It would be desirable to run a model with parameters appropriate to the TMC.
It would also be desirable to observe more sources behind the TMC and other molecular clouds.

\subsection{\hh\ ortho:para, gas T and b, and $X_{CO}$}

A determination of the \hh\ ortho:para ratio in cold molecular clouds would be of interest.
We did not detect S(1) (ortho-\hh ) absorption.
Our limit is not strong, but toward the two sources with the strongest S(0) absorption, Elias 3-16 and Tamura 8, we can rule out S(1) absorption stronger than about one half of the S(0) absorption.
The S(1) absorption per molecule in the J=1 state is 0.58 times the S(0) absorption per molecule in the J=0 state.
Consequently, the ortho:para ratio must be less than one, consistent with the low temperature value of zero, and clearly less than the high temperature value of three.

The interstellar gas temperature and line width may also be of interest.
Toward the sources both behind and embedded in the TMC the CO temperature lies between 8.4\,K and 12.0\,K, with an average value of 10.0\,K.
Toward these sources the CO Doppler b, or $e^{-1}$, linewidth lies between 0.33\,\kms\ and 1.5\,\kms , with an average value of 0.58\,\kms .
The variation among these sources of T is probably not significant, but the variation of b probably is.
It may be due to multiple velocity components along the lines of sight or turbulence.

Finally, we note that the value of $\rm X_{CO} = N_{H_2} / I_{CO}$, which is often used to interpret molecular abundances and cloud masses, is not necessarily affected by a change in \hhco\.
Most determinations of $\rm X_{CO}$ are made by comparing the integrated intensity of the $^{12}$CO J=1-0 line to \hh\ column densities determined from extinction measurements, with an assumed \hhav\ ratio.
Thus they do not depend directly on \hhco , and our measured \hhav\ ratio is consistent with that typically assumed.
\citet{lee14} discuss the cause of the variation in $\rm X_{CO}$ in the Perseus molecular cloud, and conclude that the variations are primarily caused by variations in density, turbulence, and the interstellar radiation field.

\acknowledgments
We thank M.K. Mulrey for assistance with the observations, the IGRINS team for their support of IGRINS, the McDonald Observatory staff for their support of the 2.7m telescope, and Jorge Pineda for providing their values of \nco\ and \av\ along our lines of sight.
This work used the Immersion Grating Infrared Spectrograph (IGRINS), which was developed under a collaboration between the University of Texas at Austin and the Korea Astronomy and Space Science Institute (KASI) with the financial support of the US National Science Foundation under grant AST-1229522, of the University of Texas at Austin, and of the Korean GMT Project of KASI.
The study has been supported in part by NSF grants AST-1211585 and AST-1616040
to C.S.
Our research has made use of the SIMBAD database, operated at CDS, 
Strasbourg, France.

\end{document}